\documentclass[final, 5p,times,twocolumn,authoryear]{elsarticle}

\usepackage{amssymb}
\usepackage{lipsum}
\usepackage{aas_macros}
\usepackage{xcolor}
\usepackage{amsmath}

\usepackage{lineno}
\usepackage{subfig}
\usepackage{url}

\journal{High Energy Astrophysics}

\begin{document}

\begin{frontmatter}

\title{BL Lacertae under the Flare of 2024: Probing Temporal and Spectral Dynamics}

\author[affil]{Joysankar Majumdar}
\author[affil]{Sakshi Maurya}
\author[affil]{Raj Prince\corref{cor1}}
\ead{priraj@bhu.ac.in}
\cortext[cor1]{Corresponding author}
\affiliation[affil]{organization={Department of Physics, Institute of Science, Banaras Hindu University},
            city={Varanasi},
            postcode={221005}, 
            state={Uttar Pradesh},
            country={India}}

\begin{abstract}
In October 2024, the object BL Lacertae experienced the brightest flaring event in gamma-ray ($>$100 MeV) with a historically bright $\gamma$-ray flux of $\sim$2.59 $\times 10^{-5}$ erg cm$^{-2}$ s$^{-1}$ with a detection of a 175.7 GeV photon with Fermi-LAT. This event was also followed by very high-energy $\gamma$-ray detection with LHAASO, VERITAS, and MAGIC. Soon after, Swift-XRT and Swift-UVOT follow-up confirmed the concurrent flare in X-ray, UV, and optical bands. A minimum flux doubling/halving time of 1.06 $\pm$ 0.26 hour with 4$\sigma$ significance has been observed with the Fermi-LAT orbit binned light curve. No compelling correlation has been found between $\gamma$-ray spectral indices and fluxes. The log-normal $\gamma$-ray flux distribution during the flare confirms the multiplicative nature of the non-linear perturbation causing the flare. We applied a one-zone leptohadronic model to fit the broadband SED during the flaring period. The broadband SED modeling reveals that the sudden enhancement of the magnetic field and bulk factor might promote the flare. The SED modeling also suggested a more compact emission region, which may be described by a shorter variability time than the observed one. The hadronic part best fitted the high energy part of the spectrum, suggesting the jets of BL Lac could provide a promising environment to accelerate the cosmic ray particles, such as protons. The jets of BL Lacertae could also be the possible source of astrophysical neutrinos, as an upper limit on neutrinos has already been reported from IceCube. 
\end{abstract}


\end{frontmatter}

\section{Introduction} \label{sec:intro}
Blazars are an interesting type of active galactic nuclei (AGNs) that have highly relativistic jets along the line of sight (angles less than $\sim14^\circ$) to the observer \citep{1995PASP..107..803U}. Massive black holes power the highly luminous and powerful jets of blazars, and they dominate the extragalactic $\gamma$-ray sky. Blazars are further classified into two groups- Flat-spectrum radio quasars (FSRQ) and BL Lac objects (BL Lacs). The radiation emitted from the jets expands over the whole electromagnetic spectrum \citep{1978PhyS...17..265B}. Abrupt flaring episodes are observed across different wavelengths with wide-ranging timescales from minutes to years \citep{2020NatCo..11.4176S,2021MNRAS.502.5245P}. The generation of observed emissions and properties of the emission region are still uncertain. Temporal variability study and spectral energy distribution (SED) modeling provide crucial insights into the physical properties of jets in blazars.
\begin{figure*}[ht]
    \centering
    \includegraphics[width=0.9\linewidth]{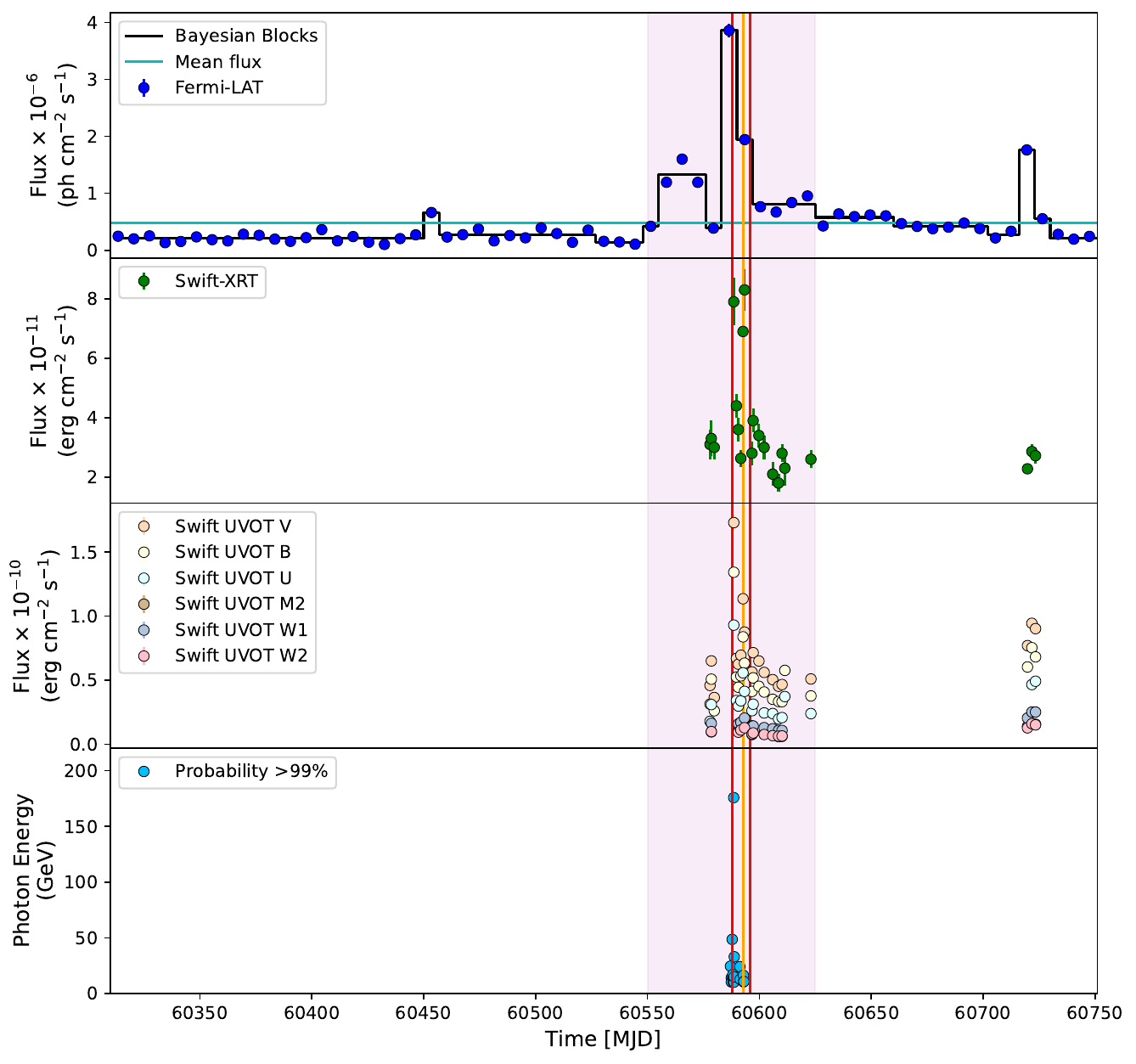}
    \caption{Multi-waveband light curve of BL Lacertae from January 2024 to March 2025. The red vertical line represents the VHE detection by VERITAS and LHAASO. The orange vertical line represents the VHE detection by MAGIC, and the brown vertical line represents the NuSTAR observation.}
    \label{fig:multi_bands_lc}
\end{figure*}

\begin{figure*}[ht]
    \centering
    \includegraphics[width=1\linewidth]{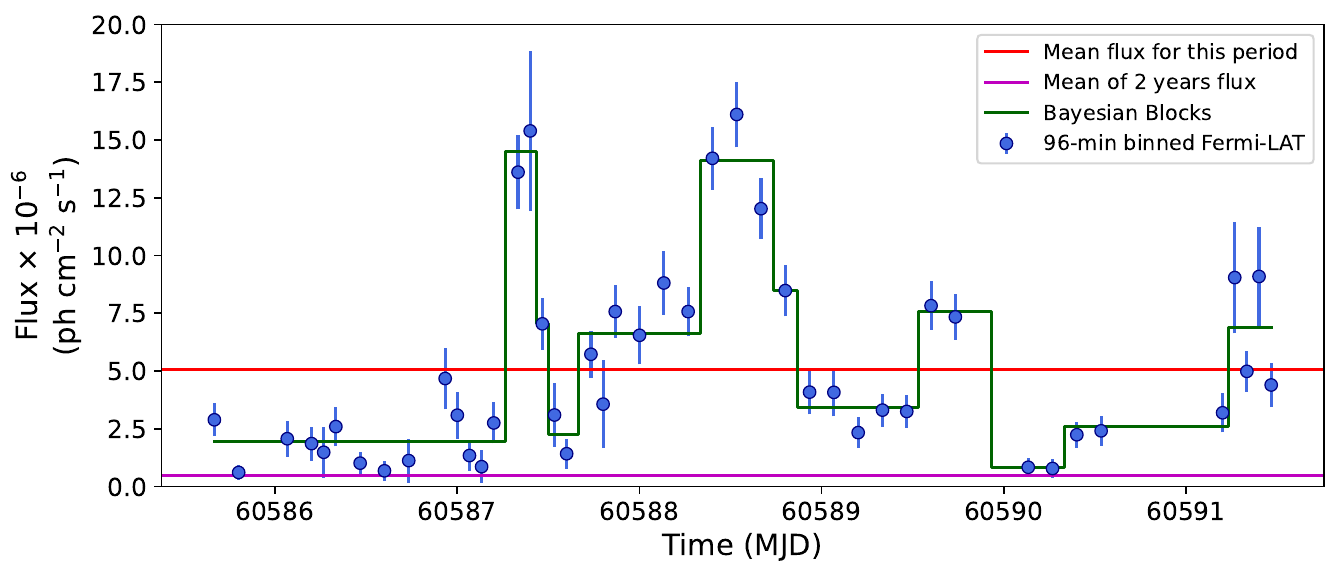}
    \caption{Fermi-LAT orbit-binned light curve (96 min) from MJD 60585.5 to MJD 60591.5.}
    \label{fig:flare_95lc}
\end{figure*}

\begin{figure*}[ht]
    \centering
    \subfloat[]{\includegraphics[width=0.48\textwidth]{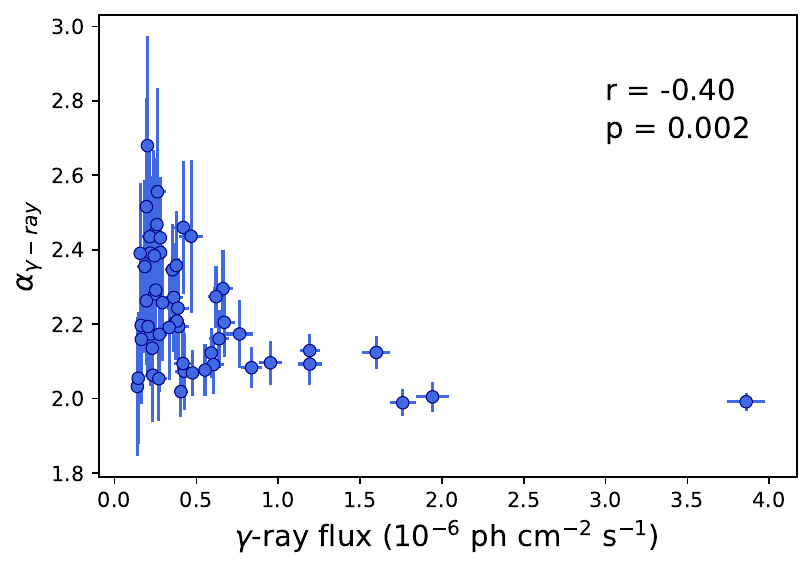}}
    \hfill
    \subfloat[]{\includegraphics[width=0.49\textwidth]{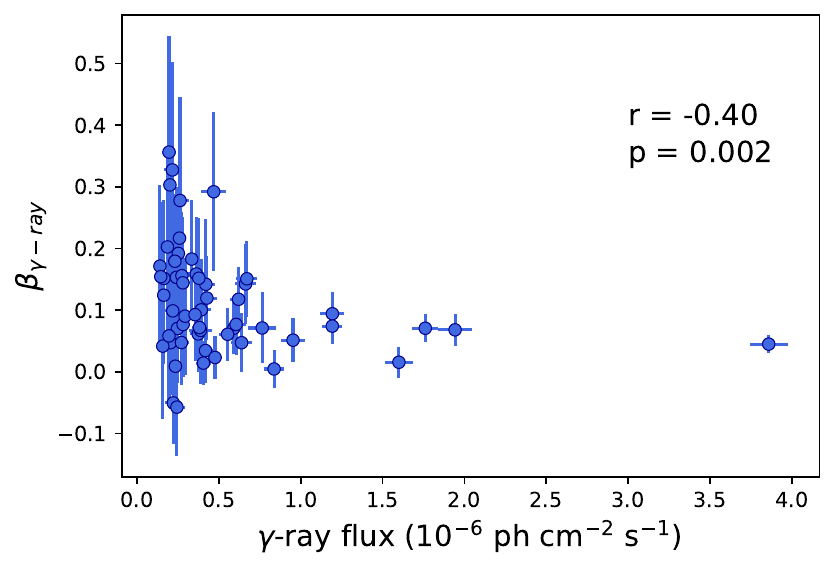}}
    \caption{(a) Variation of $\gamma$-ray flux with the spectral index of the log-parabola model used to fit each of the $\gamma$-ray light curve bins. (b) Variation of $\gamma$-ray flux with the curvature of the log-parabola model used to fit each of the $\gamma$-ray light curve bins.}
    \label{fig:index_flux}
\end{figure*}

Two evident humps are observed in the broad-band SED of blazars, which is ascribed to radiative losses experienced by a non-thermal electron distribution \citep{2010ApJ...716...30A}. The low-energy hump is from synchrotron radiation emitted by a population of relativistic electrons in the jet as they lose energy in a magnetic field. The high-energy hump is from the inverse Compton (IC) process and the photons involved in the IC process may come from internal to the jet (synchrotron self-Compton; \citealt{2009ApJ...704...38S}) or external radiation fields (external Compton; \citealt{1992A&A...256L..27D,1994ApJ...421..153S}). On the other hand, the hadronic processes involving protons can also be used to explain the second hump in the SED. 
One of the principal hadronic processes that produces $\gamma$-rays is inelastic proton-proton interaction with subsequent decay of $\pi^\circ$ and $\eta$-mesons into $\gamma$-rays \citep{2006PhRvD..74c4018K}. Thus, in hadronic models, accelerated protons are present in the jet along with electrons, and the SED's high-energy part is assumed to result from proton-initiated processes.

\citet{2020A&A...634A..80R} observed majority of blazars in Fermi-LAT 3FGL catalog show a variability of 20 days. Minimum $\gamma$-ray variability of 1.34 $\pm$ 0.30 days and X-ray variability of 3.24 $\pm$ 2.65 days have been observed in PKS 0346-27 during 2019 to 2021 \citep{2023MNRAS.520.2024K}. Even shorter time scale variability has been observed in blazars as \citet{2014Sci...346.1080A} found about 4.8 minutes $\gamma$-ray variability in IC 310 \citet{2018ApJ...854L..26S} found $\sim$5 minutes $\gamma$-ray variability in CTA 102 and \citet{2020NatCo..11.4176S} found $\sim$ 8 minutes $\gamma$-ray variability in 3C 279. Thus, variability in $\gamma$-ray helps to estimate the emission region size and location in the jet. SED modeling reveals that the location of the emission region of OQ 334 was inside the broad line region (BLR) during the flare of 2018 \citep{2021MNRAS.502.5245P}. VHE $\gamma$-ray was detected from S3 1227+25 with a shortest variability time of 6.2$\pm$0.9 hr, and the corresponding broadband SED was described by a simple one-zone leptonic SSC radiation model \citep{2023ApJ...950..152A}.

BL Lacertae is the archetype of BL Lacs with a redshift of $z = 0.0668 \pm 0.0002$ \citep{1977ApJ...212...94H}. \citet{2011ApJ...730..101A} conducted the first multi-wavelength study of BL Lacertae and observed a low activity state of BL Lacertae from August to October 2008 over all the wavebands and mild $\gamma$-ray flares in April 2009 and January 2010 with no correlation with other wavebands. They showed the presence of BLR in BL Lacertae, as suggested by broadband spectral energy distribution (SED) modeling, though BL Lacertae is classified as a BL Lac object. Multi-wavelength variability in 2019 was also reported by \citet{2020ApJ...900..137W} and they also observed the shortest optical variability time of $\sim$0.5 hr with Transiting Exoplanet Survey Satellite (TESS). 13 years (2008 August to 2021 March) of BL Lacertae's multi-wavelength data showed variability at all wavelengths and highest $\gamma$-ray flux of 4.39 $\pm$ 1.01 $\times$ 10$^{-6}$ ph cm$^{-2}$ s$^{-1}$ was detected in this period by \citet{2022MNRAS.513.4645S}. During the $\gamma$-ray flare from August to October 2020, \citet{2021MNRAS.507.5602P} observed $\gamma$-ray variability time of 8.34 hr and X-ray variability time of 11.28 hr. Even broadband SED modeling of both flaring and quiet states suggested the presence of BLR and dusty torus (DT) in BL Lacertae (\citealt{2021MNRAS.507.5602P}). \citet{2022MNRAS.509...52D} analyzed the simultaneous X-ray observations and reported a fast X-ray variability time of 60 s. \citet{2023MNRAS.521L..53A} also observed the $\gamma$-ray variability of BL Lacertae from 2020 to 2021 and tried to explain the variability using magnetic reconnection. On 27 April 2021, enhanced $\gamma$-ray activity from BL Lacertae was observed with Fermi-LAT (ATel\#14583; \citealt{2021ATel14583....1L}). With a 2-min binned $\gamma$-ray light curve, \citet{2022A&A...668A.152P} reported nearly 1-minute $\gamma$-ray variability on 27 April 2021, which was the shortest GeV variability timescale ever observed from blazars. Recently, \citet{2025MNRAS.536.1251W} analyzed multiple flares in multi-wavelengths from 2022 to 2023 and suggested a positive correlation between $\gamma$-ray and optical. Thus, BL Lacertae is a variable blazar, and variability of different timescales has been observed in $\gamma$-ray data.

In October 2024, Fermi-LAT reported enhanced $\gamma$-ray activity in BL Lacertae (ATel\#16849; \citealt{2024ATel16849....1V}). Following this Fermi-LAT detection, VERITAS detected $\gamma$-ray flaring (ATel\#16854; \citealt{2024ATel16854....1C}), MAGIC detected very-high-energy $\gamma$-ray (ATel\#16861; \citealt{2024ATel16861....1P}), LHAASO detected $\gamma$-ray flare above 1 TeV (ATel\#16850; \citealt{2024ATel16850....1X}) and high activity in optical band detected in DFOT (Atel\#16856; \citealt{2024ATel16856....1K}) and LAST (ATel\#16865; \citealt{2024TNSAN.294....1G}) on October 2024. Thus, on 5 October 2024, BL Lacertae exhibited the highest $\gamma$-ray flare with a 3-day binned flux of 6.59 $\times$ 10$^{-6}$ ph cm$^{-2}$ s$^{-1}$ in the 0.1 - 100 GeV energy range. The detection of VHE $\gamma$-ray motivated us to use the lepto-hadronic model for SED modeling.

Our motivation in this paper is to identify the fast temporal flux variation in the flare of October 2024 and to understand the physical mechanism behind the flaring episode by SED modeling.

\section{Observations and data analysis} \label{sec:obs_data}
\subsection{Fermi-LAT} \label{subsec:lat}
We acquired $\gamma$-ray data from \texttt{Fermi} Large Area Space Telescope (LAT) from 01 January 2024 to 17 March 2025 in the 100 MeV to 500 GeV range within a 30$^\circ$ radius around the source. We have adopted \texttt{FermiPy}\footnote{\url{https://github.com/fermiPy/fermipy}} \citep{2017ICRC...35..824W} for LAT data analysis. We considered all the SOURCE class events (evclass = 128 and evtype = 3) in the circular region of interest (ROI) centered at the source with a radius of 15$^\circ$. Standard cut of zenith angle $< 90^\circ$ and filter of \texttt{(DATA\textunderscore QUAL$>$0)\&\&(LAT\textunderscore CONFIG==1)} were applied to obtain good time intervals (GTI). Our model included all the sources from the Fermi-LAT Fourth Source catalog (\citealt{2020ApJS..247...33A}) within the ROI along with the galactic interstellar emission model (\texttt{gll\textunderscore iem\textunderscore v07}\footnote{\url{https://fermi.gsfc.nasa.gov/ssc/data/access/lat/BackgroundModels.html}}) and the isotropic diffuse emission template (\texttt{iso\textunderscore P8R3\textunderscore SOURCE\textunderscore V3\textunderscore v1}\footnote{\url{https://fermi.gsfc.nasa.gov/ssc/data/access/lat/BackgroundModels.html}}) for background modeling. After the initial fit of all the spectral parameters of all the sources, all the sources within a 3$^\circ$ radius of BL Lacertae were free for light curve production. 4 energy bins per decade were used in the configuration file during the SED generation.

We used the task \texttt{gtsrcprob}\footnote{\url{https://fermi.gsfc.nasa.gov/ssc/data/p7rep/analysis/scitools/help/gtsrcprob.txt}} to find the energy associated with the sources inside the circular region of 0.5$^\circ$ around BL Lacertae with event class of \texttt{P8R3\_ULTRACLEAN} (512).

\subsection{Swift XRT \& UVOT} \label{subsec:xrt_uvot}
Between 2024 and 2025, only during the flaring episodes of October 2024, observations from the X-ray telescope (XRT; 0.3–10.0 keV) of the Neil Gehrels Swift Observatory are available. The Swift-XRT spectra in the energy range 0.3-10 keV were retrieved from the public online tool the \texttt{Swift-XRT data products generator}\footnote{\url{https://www.swift.ac.uk/user_objects}} \citep{2009MNRAS.397.1177E}.

Swift's UVOT instrument observed the source in its three optical (v, b, and u) and three ultraviolet (uvw1, uvm2, and uvw2) filters. We followed the standard UVOT analysis methods\footnote{\url{https://www.swift.ac.uk/analysis/uvot/}} with the HEASoft package version 6.33.2 and the CALDB version 20240522. For each filter, we combined the observations over the specified period using the \texttt{UVOTIMSUM} task and determined the target's magnitude with \texttt{UVOTSOURCE}. The magnitudes were corrected for Galactic extinction following \citet{2011ApJ...737..103S}. The source magnitude was extracted using a 3.0 arcsec circular region centered on the source, while the background magnitude was obtained from a 10 arcsec circular region nearby. Finally, we converted the corrected magnitudes to fluxes using the zero points from \citet{2011AIPC.1358..373B} and the conversion factors from \citet{2016MNRAS.461.3047L}.

\subsection{NuSTAR}
We have only one NuSTAR observation on 13 October 2024 with a 24 kilo-sec exposure. We used \texttt{NuSTARDAS version 1.9.2} provided by \texttt{HEASoft} with the CALDB version 20240812 for data analysis. We used a 45 arcsec radius around the source and a 90 arcsec radius around the source to extract the spectrum of the source and background, respectively. We used the task \texttt{nupipeline} to produce cleaned event files filtered for a GTI and \texttt{nupipeline} to create the source and background spectrum. The spectrum is fitted in Xspec to extract the flux and the spectral index.

\section{Results and Discussion} \label{sec:results}

\subsection{Multi-wavelength light curve}
Multi-wavelength light curves of BL Lacertae have been shown in Figure \ref{fig:multi_bands_lc}. The first panel shows the weekly binned $\gamma$-ray light curve in the energy range of 0.1 - 500 GeV from MJD 60310 to MJD 60751. A few Swift observations for the source between MJD 60570 and MJD 60625 have been found. Swift-XRT and UVOT light curves have been shown in the second and third panels of Figure \ref{fig:multi_bands_lc}. Due to the limited number of Swift observations in the whole period, it was not possible to perform the correlation study using a discrete correlation function (DCF). However, visually, it can be made out that the fluxes across the wavebands are correlated (Figure \ref{fig:multi_bands_lc}).

We implemented the Bayesian blocks (BB) algorithm \citep{2013ApJ...764..167S} with a false-alarm probability (p$_0$) of 0.05 on the 7-day binned (first panel of Figure \ref{fig:multi_bands_lc}) $\gamma$-ray light curve with the help of a PYTHON package \texttt{lightcurves}\footnote{\url{https://github.com/swagner-astro/lightcurves}} \citep{2022icrc.confE.868W} and the BB blocks are shown with blue lines. The BB blocks helped to identify the flaring state in the light curve. Along with the BB blocks, we applied the HOP algorithm \citep{1998ApJ...498..137E} to identify the flare period. For flare detection using the HOP algorithm, we applied the condition of BB block flux F$_{BB} \geq 3\bar{F}$, where $\bar{F}$ is the average flux of the 7-day binned light curve. Thus, we have selected MJD 60550 to MJD 60625 as our main flaring period, indicated in the purple shaded region in Figure~\ref{fig:multi_bands_lc}.

We calculated the energy associated with the photons during the flaring period, and the photons detected with a 99\% detection probability above 10 GeV can be seen in the last panel of Figure \ref{fig:multi_bands_lc}. We detected the highest energetic photon of 175.7 GeV in Fermi-LAT on 5 October 2024 from BL Lacertae with the probability of detection more than 99.99\%, but the historically highest energy photon ever detected from BL Lacertae with Fermi-LAT was 238 GeV, reported by \citet{2021MNRAS.507.5602P}.

The highest energy photon detected by LAT was also supported by other very high-energy (VHE) telescopes. VERITAS detected VHE $\gamma$-ray above 200 GeV with a flux of about 20\% of the Crab Nebula flux on 5 October 2024. An intra-night light curve of BL Lacertae above 200 GeV showed a sub-hour variability on 5 October 2024 \citep{2024ATel16854....1C}. LHAASO also detected VHE $\gamma$-ray above 1 TeV of flux around 50\% of Crab Nebula flux on 5 October 2024 \citep{2024ATel16850....1X}. The VERITAS and LHAASO detection is marked with a red vertical line in Figure \ref{fig:multi_bands_lc}. MAGIC telescopes also detected VHE $\gamma$-ray flux comparable to Crab Nebula flux on 10 October 2024 above 250 GeV, and they also found that the flux changed by a factor of 2 within an hour \citep{2024ATel16861....1P}. This MAGIC detection has been shown in Figure \ref{fig:multi_bands_lc} with an orange vertical line. We found only a single NuSTAR observation on 13 October 2024, which is indicated by a brown line in Figure \ref{fig:multi_bands_lc}.

\subsection{$\gamma$-ray temporal and spectral variability}\label{subsec:gamma_Var}
To further investigate the fast variability, we produced LAT orbit-binned light curves from MJD 60585.5 to MJD 60591.5 and applied the BB algorithm to identify the flux changes as shown in Figure \ref{fig:flare_95lc}. We have estimated the shortest flux doubling/halving timescale from the orbit-binned light curve, using the formula as follows \citep{2011A&A...530A..77F}
\begin{equation} \label{eq:flux_doubling_time}
    F(t_2) = F(t_1) \times 2^{\frac{\Delta t}{t_d}}
\end{equation}
where $F(t_1)$ and $F(t_2)$ are the flux values at times $t_1$ and $t_2$ , respectively, $\Delta t = t_1 - t_2$ , and $t_d$ denotes the flux doubling/halving timescale. We found the shortest significant flux doubling time $t_d = 1.06 \pm 0.26$ hour with 4$\sigma$ significance. Assuming the emission region to be a spherical blob, we estimated the emission region size using
\begin{equation}
    R \leq c\ t_d \frac{\delta}{1+z},
\end{equation}
where c is the speed of light in vacuum, $t_d$ is the minimum flux doubling/halving time, $\delta$ is the Doppler factor, and z is the redshift. Thus, using $\delta$ = 11.55 \citep{2020ApJ...897...10Z} and $t_d \sim$ 1.06 hour, we found R $\leq$ 1.2 $\times$ 10$^{15}$ cm. We tried to explore the minute-scale variability using 5-min, 3-min, and 2-min binned light curves, but no significant variability was found.

To study variability, we have calculated the fractional root mean square (rms) variability amplitude (F$_{var}$) as described in \citet{2003MNRAS.345.1271V}. Using the 7-day binned $\gamma$-ray light curve, we found F$_{var}$ = 1.19 $\pm$ 0.01, indicating variability above 100 per cent and similarly \citet{2021MNRAS.507.5602P} found F$_{var}$ = 1.84 $\pm$ 0.02 during 2020.

We also derived the correlation between $\gamma$-ray flux and the spectral index ($\alpha$) and curvature ($\beta$) as shown in Figure \ref{fig:index_flux}. A mild harder-when-brighter trend is seen. The Spearman’s rank correlation analysis between $\gamma$-ray flux and $\alpha_{\gamma-ray}$ and $\beta_{\gamma-ray}$ both resulted in a rank coefficient of $r=-0.40$, with a null-hypothesis probability of $P=0.002$. The $\alpha_{\gamma-ray}$ and $\beta_{\gamma-ray}$ vary similarly with flux.

\begin{figure}[ht]
    \centering
    \includegraphics[width=0.95\linewidth]{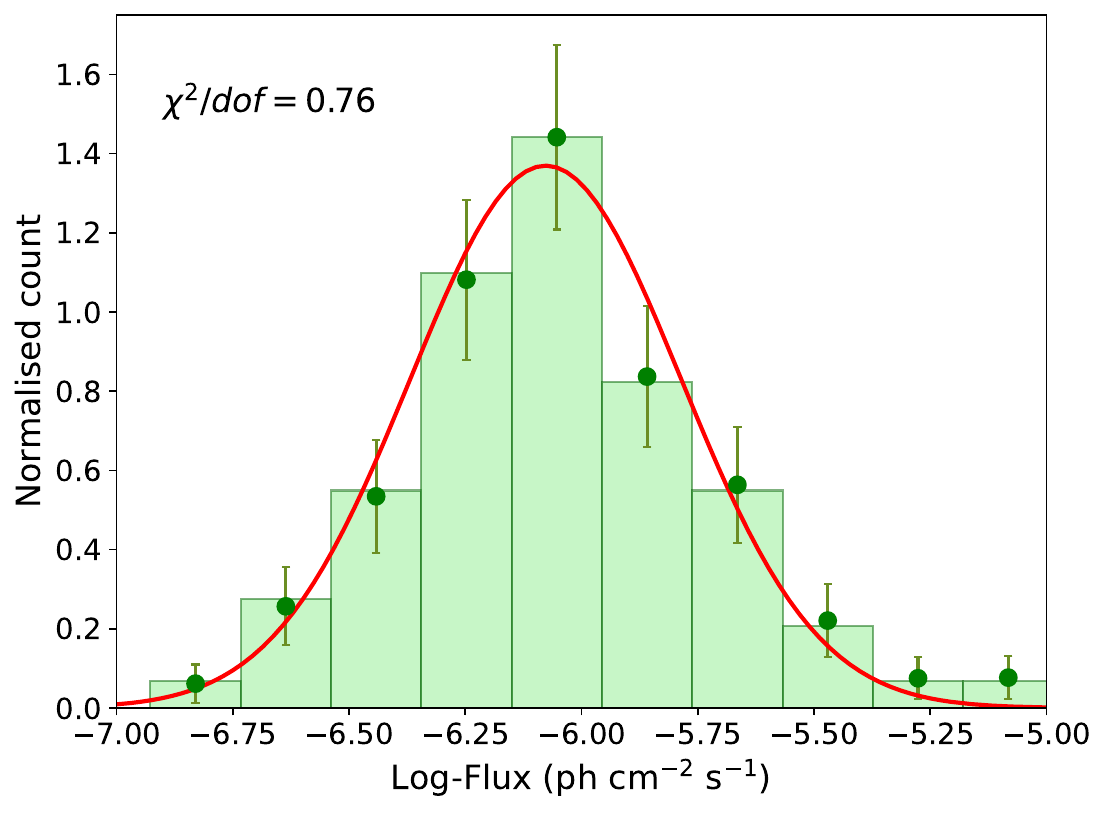}
    \caption{$\gamma$-ray flux distribution during flaring state.}
    \label{fig:flux-dist}
\end{figure}

We studied $\gamma$-ray flux distribution. The flux distribution helps us to understand the nature of variability, whether it is caused by additive or multiplicative processes in the jet. A Gaussian flux distribution suggests an additive process caused by stochastic and linear variations \citep{2010LNP...794..203M}, but a log-normal distribution suggests non-linear stochastic variation \citep{2010LNP...794..203M,2005MNRAS.359..345U}. Mostly, a log-normal flux distribution is observed in AGNs \citep{2018RAA....18..141S,2021MNRAS.502.5245P,2024MNRAS.527.2672S}. In this study, we have produced the flux distribution during the flaring state with the 1-day binned light curve, as shown in Figure \ref{fig:flux-dist}. We fitted a log-normal distribution with a mean of -6.08 $\pm$ 0.6 and a standard deviation of 0.29 $\pm$ 0.02, as shown with red color in Figure \ref{fig:flux-dist} along with reduced $\chi^2$ of 0.76. Thus, the log-normal flux distribution suggested that the multiplicative nature of the non-linear perturbation caused the flaring event.

\begin{figure}[ht]
    \centering
    \includegraphics[width=0.99\linewidth]{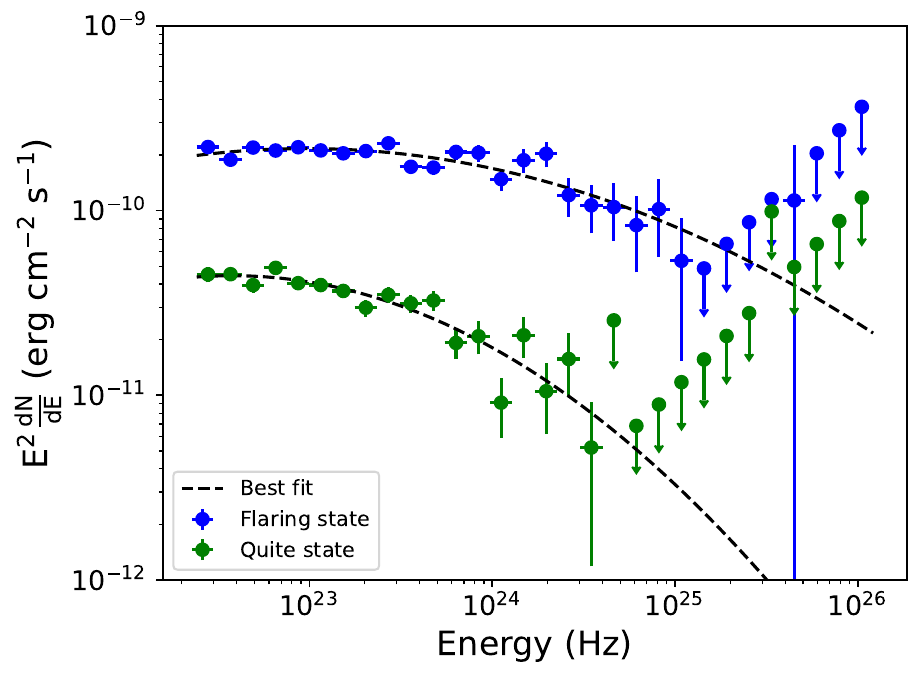}
    \caption{$\gamma$-ray SEDs for the flaring and quiet states.}
    \label{fig:gamma-sed}
\end{figure}

We produced the $\gamma$-ray SEDs for the flaring period and a quiet period from MJD 60310 to MJD 60500 as shown in Figure \ref{fig:gamma-sed} with blue and green points, respectively. Clearly, it is visible that the flux density in the flaring state is much higher than the quiet state in Figure \ref{fig:gamma-sed}. The SEDs are best fitted with a log-parabola model as shown with a black dotted line in Figure \ref{fig:gamma-sed}. Log-parabola model in flaring state yields an index ($\alpha$) of 2.08 $\pm$ 0.02 and a curvature ($\beta$) of 0.05 $\pm$ 0.009 and the quite state yields $\alpha = 2.31 \pm 0.03$ and $\beta = 0.08 \pm 0.02$. A break or curvature above 10 GeV has been observed in both the $\gamma$-ray SEDs, indicating photons above 10 GeV are getting absorbed. Thus, the curvature in $\gamma$-ray spectra can be considered as a signature of photon-photon absorption (pair-production), where a $\gamma$-ray photon interacts with low-energy photons from the BLR, suggesting the emission region is possibly within the BLR \citep{2006ApJ...653.1089L}. Being a BL Lac type object, BL Lacertae should not have a standard BLR, but $\gamma$-ray spectra forced to consider the existence of BLR, and thus, further we have taken BLR into consideration during broadband SED modeling.

\subsection{X-ray temporal and spectral variability}
We calculated the flux doubling or halving time within the X-ray light curve shown in the second panel of Figure \ref{fig:multi_bands_lc}. The minimum flux doubling time is found to be 0.71 $\pm$ 0.07 day. On MJD 60591, the observed X-ray flux was (2.63 $\pm$ 0.28) $\times$ 10$^{-11}$ erg cm$^{-2}$ s$^{-1}$, on MJD 60592 flux was (6.90 $\pm$ 0.50) $\times$ 10$^{-11}$ erg cm$^{-2}$ s$^{-1}$ and on MJD 60593, flux was (8.30 $\pm$ 0.90) $\times$ 10$^{-11}$ erg cm$^{-2}$ s$^{-1}$. These consecutive flux enhancements suggest a fast X-ray flux variability within a day. Again, this suggests that a compact emission region is involved in the production of such a high variability event.

\begin{figure}[ht]
    \centering
    \includegraphics[width=0.95\linewidth]{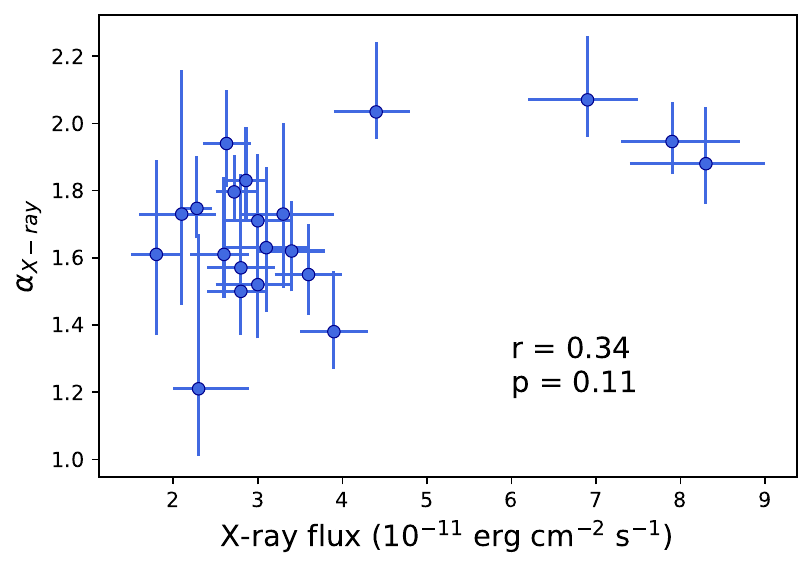}
    \caption{Flux vs PL index for X-ray observations}
    \label{fig:xray_flux_index}
\end{figure}

We have seen a low positive correlation between X-ray flux and PL index ($\alpha_{X-ray}$) with Spearman’s rank correlation coefficient $r = 0.34$ and the corresponding p-value = 0.11. As shown in Figure~\ref{fig:xray_flux_index}, we can see a mild \textit{softer-when-brighter} trend in X-rays.

\begin{figure}[ht]
    \centering
    \includegraphics[width=0.95\linewidth]{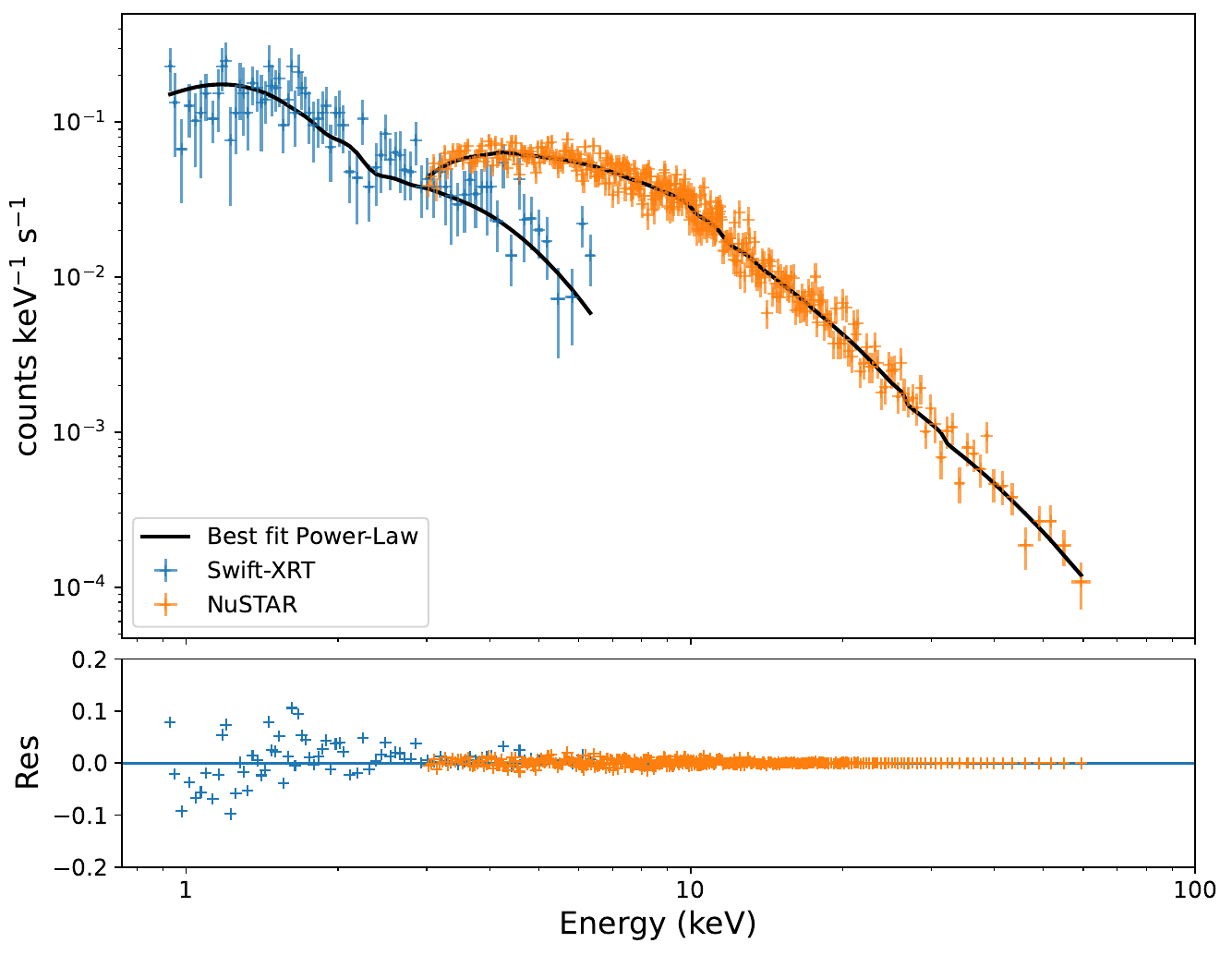}
    \caption{Combined Swift-XRT and NuSTAR spectra fitted with a simple power law model. The lower panel shows the residual of the fit.}
    \label{fig:xray_spectra_fit}
\end{figure}

We combined the Swift-XRT and NuSTAR spectra and performed a combined power-law fit as shown in Figure \ref{fig:xray_spectra_fit}. We have used the \texttt{constant$\times$tbabs$\times$powerlaw} model in \texttt{xspec} for fitting the data. While the \texttt{constant} for the Swift-XRT spectra was kept fixed to 1 and the \texttt{constant} for the NuSTAR spectra was kept free during the fit, which resulted in the constant being 1.39. For the \texttt{tbabs} model, the Hydrogen Column density (nH) was kept fixed to 1.75 $\times$ 10$^{21}$ cm$^{-2}$ \citep{2016A&A...594A.116H}. We found the combined power law index to be 1.66 $\pm$ 0.03. The fit was executed with a $\chi^2_{red} = 0.99$. This suggests that a single power-law can explain the X-ray spectrum from soft to hard X-rays, proving a non-thermal origin for them.

\subsection{Broadband SED modeling}
\begin{figure*}[ht]
    \centering
    \includegraphics[width=0.8\linewidth]{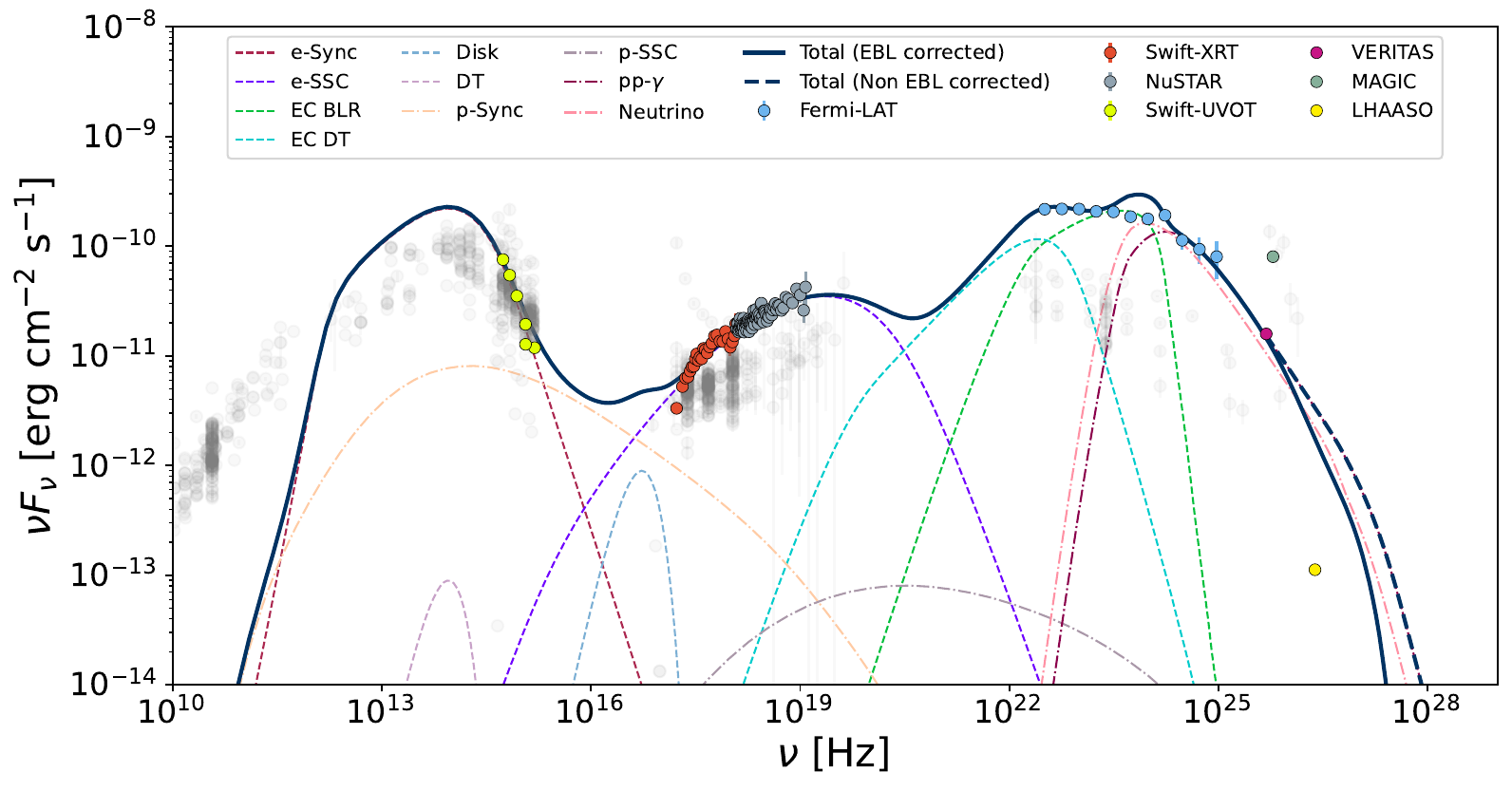}
    \caption{Broadband SED of the flaring state. The grey data points in the background represent the archival average SED state extracted from the SED builder. }
    \label{fig:flare_broadband_sed}
\end{figure*}

The broadband SED for the flaring state was curated from VERITAS, LAHAASO, MAGIC, Fermi-LAT, Swift-XRT, Swift-UVOT, and NuSTAR data, as shown in Figure \ref{fig:flare_broadband_sed}. 
A preliminary analysis from VERITAS suggested that the flux of BL Lacertae drops from 20\% of Crab Nebula flux to 10\% of Crab Nebula flux above 200 GeV from 5 October 2024 to 7 October 2024 \citep{2024ATel16854....1C}. The constant VHE flux above 100 GeV from the Crab is about (4.95$\pm$0.03) $\times 10^{-10}$ ph cm$^{-2}$ s$^{-1}$ \citep{2023ApJ...956...80A}, and we used this flux to produce the SED point for VERITAS. The MAGIC collaboration also reported a maximum flux of 2$\times 10^{-10}$ ph cm$^{-2}$ s$^{-1}$ above 250 GeV on 10 October 2024 \citep{2024ATel16861....1P}. The LHAASO collaboration also reported the detection of 0.5 Crab unit flux above 1 TeV on 5 October 2024. \citet{2024ApJ...971L..45C} observed the flux 7.0 $\pm$ 1.1 ph cm$^{-2}$ s$^{-1}$ for NGC 4278 in the 1-10 TeV energy range by LHAASO, which is approximately 5\% of the Crab Nebula. We used the value to convert the LHAASO-detected flux of BL Lacertae for the broadband SED. We used publicly available SED modeling code \texttt{JetSeT}\footnote{\url{https://github.com/andreatramacere/jetset}} \citep{2020ascl.soft09001T,2011ApJ...739...66T,2009A&A...501..879T} to perform the lepto-hadronic modeling of the constructed broadband SED. 
We considered a one-zone (spherical blob) lepto-hadronic case for the SED modeling. We only considered the proton-proton interaction for the hadronic case, as \texttt{JetSeT} only provides p-p interaction based on \citet{2006PhRvD..74c4018K}. Electron and proton populations described by a broken power law (bkn) having a low energy slope of p$_1$, high energy slope p$_2$, and break energy of $\gamma_{break}$ as described by \citet{2009A&A...501..879T} have been considered for this study. The broadband emission from the jet depends on the radius of the emission region (R), the magnetic field (B), the bulk Lorentz factor ($\Gamma$), and the viewing angle ($\theta$) of the jet. The modeling results are tabulated in Table \ref{tab:sed_params} and depicted in Figure \ref{fig:flare_broadband_sed}.

The p-p interactions have been used to estimate the high-energy $\gamma$-ray part as they have a parameter space to interpret the $\gamma$-ray spectra \citep{2022A&A...659A.184L} and they can explain the VHE spectra \citep{2022PhRvD.106j3021X}. The p-p
interactions are comprised of the following reactions \citep{2025ApJ...980...19O}
\begin{align} 
p + p &=  p \\ 
p + p &= \pi^\circ \longrightarrow \gamma + \gamma \\
p + p &= \pi^+ \longrightarrow \mu^+ + \nu_\mu \longrightarrow e^+ + \nu_e +\bar{\nu}_\mu + \nu_\mu \\
p + p &= \pi^- \longrightarrow \mu^- + \bar{\nu}_\mu \longrightarrow e^- + \bar{\nu}_e + \nu_\mu + \bar{\nu}_\mu
\end{align}
where $\pi^\circ$ is neutral pion, $\pi^{\pm}$ are charged pions, $\mu^{\pm}$ are charged muons, $e^{\pm}$ are positron and electron, and $\nu$ is neutrino.

Previously, the SED modeling of BL Lacertae in different epochs suggested the existence of BLR and DT components \citep{2024MNRAS.527.5140S,2021MNRAS.507.5602P}. Thus, in this present study, we also considered both BLR and DT for SED modeling. the disk luminosity (L$_{Disk}$) of the source is taken as 3.3 $\times 10^{43}$ erg s$^{-1}$ as reported by \citet{2018ApJS..235...39C}. We have fixed the Thomson depth of BLR ($\tau_{BLR}$) and DT ($\tau_{DT}$) with a standard value of 0.1 \citep{1981MNRAS.195..437S,2015MNRAS.448.1060G}. We fixed the radii of BLR with respect to disk luminosity using the relations from \citet{2007ApJ...659..997K}, as

\begin{equation}
    R_{BLR-in} = 10^{17} \times \sqrt{\frac{L_{Disk}}{10^{45}}},
\end{equation}
\begin{equation}
    R_{BLR-out} = 1.1 \times R_{BLR-in}.
\end{equation}

We have taken the initial radius of DT as 1 $\times 10 ^{17}$ cm and freed it during fitting. We have fixed the viewing angle ($\theta$) to be 0.1 \citep{2024MNRAS.527.5140S}. The rest of the parameters, like emission region height (R$_H$), temperature of the disk (T$_{Disk}$), temperature of the DT (T$_{DT}$), and emitter density (N), were kept free during the fitting process. We assumed the cold proton to relativistic electron ratio NH$_{cold\_to\_rel\_e}$ = 0.1 and kept it frozen during the fit. All the spectral parameters of the electron and proton distributions were kept free during the fitting.

Our SED modeling suggested the emission region size to be 8 $\times$ 10$^{14}$ cm, which is more compact than the emission region size predicted by \citet{2021MNRAS.507.5602P}. Even the emission region is smaller than the emission region calculated in section \ref{subsec:gamma_Var}, suggesting a faster time scale is involved in the flaring process. We found the location of the emission region to be 1.84 $\times$ 10$^{16}$ cm, which indicated the emission region is located inside the BLR, but \citep{2021MNRAS.507.5602P} found the emission region was beyond the BLR. Our SED modeling suggested a magnetic field of 4.24 G, which is much higher than both the flaring and quiet states of 2021-2022 \citep{2024MNRAS.527.5140S} and even higher than that flaring state of 2020 \citep{2021MNRAS.507.5602P}. The bulk factor was found to be 14.32 and 6.01 for two different low states in 2021-22 \citep{2024MNRAS.527.5140S}, and in our study, we found the bulk factor to be 14.11, which is comparable to one of the low states. Thus, we conclude that an increment in the magnetic field and bulk factor might have caused the flare in BL Lacertae. In this case, the injected electron index (p$_{e,1}$) of 1.65 is harder compared to the quiet states of \citet{2021MNRAS.507.5602P} and \citet{2024MNRAS.527.5140S}. The harder p$_{e,1}$ suggested more efficient particle acceleration with less radiative cooling. For the hadronic case, we only considered the proton-proton interaction, and the $\gamma$-ray and neutrino produced from this interaction are shown in Figure \ref{fig:flare_broadband_sed}. The initial part of the high energy hump in SED is explained by the external Compton by BLR and DT while the latter high energy part is well fitted by the p-p $\gamma$-ray as shown in Figure \ref{fig:flare_broadband_sed}. As VHE part of the SED is mainly modeled with the hadronic code, we applied the extragalactic background light (EBL) attenuation model from \citet{2008A&A...487..837F} to the hadron part, thus non-EBL corrected and EBL corrected both are shown in Figure \ref{fig:flare_broadband_sed}. The energy densities of every component and the luminosities of electron, magnetic field, and proton are tabulated in Table \ref{tab:sed_params}.

The total energy budget is estimated using the jet luminosities due to electrons and protons, which are 1.54 $\times$ 10$^{43}$ erg s$^{-1}$ and 2.39 $\times$ 10$^{47}$ erg s$^{-1}$, respectively. These luminosities are much higher than the jet luminosity due to the magnetic field (8.56 $\times$ 10$^{42}$ erg s$^{-1}$). It implied that the jet is particle-dominated, mostly proton-dominated. A similar range of electron and proton luminosity using the lepto-hadronic modeling had been reported for TXS 0506+056 in 2017 by \citet{2022PhRvD.106l3005S}. We calculated the Eddington luminosity (L$_{Edd}$) by using L$_{Edd}$ $\approx$ 10$^{38}$(M/M$_\odot$) erg s$^{-1}$ $\approx$ 1.6 $\times$ 10$^{46}$ erg s$^{-1}$, where M is the central SMBH mass taken as 10$^{8.21}$ M$_\odot$ from \citet{2018ApJS..235...39C}. The total jet power is about 2.39 $\times$ 10$^{47}$ erg s$^{-1}$, which is much higher than L$_{Edd}$. \citet{2025ApJ...980...19O} have observed similar jet luminosity which is exceeding the L$_{Edd}$ in S5 0716+714. The jet luminosity may exceed L$_{Edd}$ for a highly collimated jet outflow in blazars, as jets do not interfere with the accretion flow \citep{2019NatAs...3...88G}. The p-p interactions are also expected to produce neutrinos, but till now, there is no confirmation of any neutrino detection from this source. However, following the VHE activities, IceCube did search for neutrinos and detected a time-integrated muon-neutrino flux upper limit for this source of E$^2 dN/ dE = 6.6\times10^{-2} ~GeV/cm^2$ at 90\% confidence level. Our SED modeling result concludes that a hadronic contribution must be considered to explain the high-energy part of the spectrum, and BL Lac can be considered as a possible source of high-energy cosmic rays and astrophysical neutrinos.

\begin{table*}[ht]
    \centering
    \begin{tabular}{ l l l }
    \hline
        \textbf{Symbols} & \textbf{Description} & \textbf{Values}  \\ \hline
        \multicolumn{3}{c}{\textbf{Fixed Parameters}} \\ \hline
         z & Redshift & 0.0668 \\ \hline
         $\theta$ & Viewing angle (degree) & 0.1 \\ \hline
         R$_{BLR-in}$ & BLR inner radius (cm) & 1.82 $\times 10^{16}$ \\ \hline
         R$_{BLR-out}$ & BLR outer radius (cm) & 2.00 $\times 10^{16}$ \\ \hline
         $\tau_{BLR}$ & Thomson depth of the BLR & 0.1 \\ \hline
         $\tau_{DT}$ & Thomson depth of the DT & 0.1 \\ \hline
         NH$_{cold\_to\_rel\_e}$ & Cold proton to electron ratio & 0.1 \\ \hline 
         L$_{Disk}$ & Luminosity of disk (erg s$^{-1}$) & 3.3 $\times 10^{43}$ \\ \hline
         \multicolumn{3}{c}{\textbf{Free Parameters}} \\ \hline
         $\gamma_{e,min}$ & Electron's low energy cutoff & 18.92 \\ \hline
         $\gamma_{e,max}$ & Electron's high energy cutoff & 2.17 $\times 10^{6}$ \\ \hline
         $N_e$ & Injection electron density (cm$^{-1}$) & 1.62 $\times 10^4$ \\ \hline
         $\gamma_{e,break}$ & Electron break energy & 5.51 $\times 10^2$ \\ \hline
         p$_{e,1}$ & Electron low energy spectral index & 1.65 \\ \hline
         p$_{e,2}$ & Electron high energy spectral index & 6.90 \\ \hline
         T$_{DT}$ & Temperature of DT (K) & 1.19 $\times 10^3$ \\ \hline
         R$_{DT}$ & Radius of DT (cm) & 9.99 $\times 10^{16}$ \\ \hline
         T$_{Disk}$ & Temperature of Disk (K) & 7.04 $\times 10^{5}$ \\ \hline
         R & Region size (cm) & 8.0 $\times 10^{14}$ \\ \hline
         R$_H$ & Region height (cm) & 1.84 $\times 10^{16}$ \\ \hline
         B & Magnetic field (G) & 4.24 \\ \hline
         $\Gamma$ & Bulk Lorentz factor & 14.11 \\ \hline
         $\gamma_{p,min}$ & Proton low energy cutoff & 4.0 \\ \hline
         $\gamma_{p,max}$ & Proton high energy cutoff & 1 $\times 10^{6}$ \\ \hline
         $N_p$ & Injecting proton density (cm$^{-1}$) & 6.7 $\times 10^{6}$ \\ \hline
         $NH_{pp}$ & Target density (cm$^{-1}$) & 3.9 $\times 10^{6}$ \\ \hline
         $\gamma_{p,break}$ & Proton break energy & 1 $\times 10^3$ \\ \hline
         p$_{p,1}$ & Proton low energy spectral index & 3.0 \\ \hline
         p$_{p,2}$ & Proton high energy spectral index & 4.5 \\ \hline
         \multicolumn{3}{c}{\textbf{Energy Densities}} \\ \hline
         U$_e$ & Energy density of electron (erg cm$^{-3}$) & 1.29 \\ \hline
         U$_B$ & Energy density of magnetic field (erg cm$^{-3}$) & 0.72 \\ \hline
         U$_{Disk}$ & Energy density of disk (erg cm$^{-3}$) & 0.12 \\ \hline
         U$_{BLR}$ & Energy density of magnetic field (erg cm$^{-3}$) & 15.26 \\ \hline
         U$_{DT}$ & Energy density of DT (erg cm$^{-3}$) & 0.35 \\ \hline
         U$_{p}$ & Energy density of proton (erg cm$^{-3}$) & 2 $\times 10^4$ \\ \hline
         \multicolumn{3}{c}{\textbf{Jet Luminosities}} \\ \hline
         L$_e$ & Jet luminosity of electron (erg s$^{-1}$) & 1.54 $\times 10^{43}$ \\ \hline
         L$_B$ & Jet luminosity of magnetic field (erg s$^{-1}$) & 8.56 $\times 10^{42}$ \\ \hline
         L$_{p}$ & Jet luminosity of proton (erg s$^{-1}$) & 2.39 $\times 10^{47}$ \\ \hline
    \end{tabular}
    \caption{Best fit parameters of broadband SED Lepto-Hadronic modeling.}
    \label{tab:sed_params}
\end{table*}

\section{Summary}
We have applied the Bayesian blocks (BB) algorithm along with the HOP algorithm to determine the flaring state in the $\gamma$-ray light curve. Simultaneous Swift observation also confirmed the flaring state in X-ray, UV, and optical bands. A minimum flux doubling/halving time of 1.06 $\pm$ 0.26 hour in the Fermi-LAT orbit-binned light curve was observed during the flaring period, suggesting an emission region size of 1.2 $\times 10^{15}$ cm. The $\gamma$-ray variability is probed by estimating the fractional variability amplitude F$_{var}$, which is found to be 1.19 $\pm$ 0.01. A mild hint of a harder-when-brighter trend is seen for both $\alpha_{\gamma-ray}$ and $\beta_{\gamma-ray}$ variation with $\gamma$-ray flux. We found the $\gamma$-ray flux distribution during the flare to be log-normal, indicating the multiplicative nature of non-linear perturbations in the jet. The $\gamma$-ray SED during the flare is at a higher flux compared to the quiet state, and both are best fitted with a log-parabola model. The spectrum at the quiet state shows a break almost after 10 GeV, suggesting the emission region is within the BLR. However, during the high-flux state, a high-energy photon is seen in the spectrum. We found the minimum flux doubling time in X-ray to be 0.71 $\pm$ 0.07 day. A mild softer-when-brighter trend has been observed in the X-ray flux index variation. Due to concurrent flaring observation in $\gamma$-ray, X-ray, UV, and optical, we adopted a one-zone lepto-hadronic model for broadband SED modeling. SED modeling suggested an emission region size of 8.0 $\times 10^{14}$ cm, which is smaller than the predicted emission region size from the observed variability. SED modeling also suggested that the emission region is located within the BLR. The low energy hump is dominated by synchrotron emission, the X-ray part is majorly dominated by SSC and the higher energy part is best-fitted by the combination of external-Compton from BLR and DT and the p-p $\gamma$-ray. The VHE energy part of the broadband SED is well fitted by the hadronic model. The unanticipated enhancement of the magnetic field and bulk factor might have caused the flaring event.

\section{Acknowledgments}
We thank the reviewers for their constructive suggestions, which have helped to improve the manuscript. This research makes use of the publicly available data from Fermi-LAT and NuSTAR obtained from the FSSC data server and distributed by NASA Goddard Space Flight Center (GSFC). This work made use of data supplied by the UK Swift Science Data Centre at the University of Leicester. The data, software, and web tools obtained from NASA’s High Energy Astrophysics Science Archive Research Center (HEASARC), a service of GSFC, are used in this work. The authors acknowledge the support from the BHU IoE seed grant.

\bibliographystyle{elsarticle-harv} 
\bibliography{bibgpy}

\end{document}